# Modelling a Transition-Edge Sensor X-ray Microcalorimeter Linear Array for Compton Profile Measurements and Energy Dispersive Diffraction

Daikang Yan (闫代康), Lisa M. Gades, Tejas Guruswamy, Umeshkumar M. Patel, Orlando Quaranta, and Antonino Miceli

*Abstract*—Transition-edge sensors are a type of superconducting detector that offers high energy resolution based on their sharp resistance-temperature feature in the superconducting-to-normal transition. TES X-ray microcalorimeters have typically been designed and used for spectroscopic applications. In this work, we present a design optimization for a TES X-ray microcalorimeter array for high-energy scattering and diffraction measurements. In particular, Compton scattering provides information about the electron momentum distribution, while energy dispersive diffraction provides structural information about dense engineering materials. Compton scattering and energy dispersive diffraction experiments must be conducted in the very hard X-ray regime (~ 100 keV), demanding a high X-ray stopping power in the detector; therefore, an absorber with a large heat capacity is needed in conjunction with the TES. In addition, both applications would benefit from an array composed of parallel strips. We present a design for a TES X-ray microcalorimeter optimized for such applications. In particular, we model the longitudinal position dependence due to the finite thermal diffusion time in the absorber.

*Index Terms*—Compton scattering, Energy Dispersive Diffraction, Transition-edge Sensors, Position Dependence

## I. Introduction

TRANSITION-EDGE sensors (TES) have a promising future in high-energy resolution X-ray science applications [1]. Traditionally, energy-resolving experiments are accomplished with semiconductor detectors or crystal spectrometers. The former, due to its ~ eV level photoelectron excitation energy, can only reach ~ 130 eV of resolution at 6 keV for silicon (Si) detectors and ~ 500 eV at 60 keV for cryogenic germanium (Ge) detectors, due to the statistical noise of the electron-hole pairs generation-recombination process. The latter being the option when higher energy resolution is needed, however, has a relatively slow operation speed, because its energy scan is achieved with a time-consuming rotation of the dispersive crystals. TESs exhibit close to unit quantum efficiency and energy resolution of $\Delta E$ = 22 eV at 97 keV [2] and thus are a promising alternative when it comes to high-speed operation, high-energy-resolution demands. In addition, the advances in microwave multiplexing have made it possible to read out thousands of fast TESs, enabling TES spectrometers with high throughput. In this paper, we present a design for a linear array of TES, which offers one dimensional spatial resolution. Two potential uses are energy dispersive X-ray diffraction (EDXRD) and Compton scattering measurement in the very hard X-ray regime (> 30 keV). EDXRD measures the sample's diffraction pattern at a fixed angle by using polychromatic photons and an energy dispersive detector. The detector's energy resolution limits the d-spacing that can be resolved in the sample. Compared to the conventional angle dispersive X-ray diffraction that uses a monochromatic source and needs to rotate the sample, EDXRD is especially useful for situations in which the angles cannot be scanned and has the advantage that all scattering takes place into a small solid angle. This is a very attractive feature for studies in extreme environments where the need for large X-ray windows can compromise the environmental chamber (e.g., diamond anvil cell) [3]. If an EDXRD detector has spatial resolution as well, then multiple parts of the sample could be measured at the same time, thus providing simultaneous imaging and diffraction information [4]. Compton scattering is the inelastic scattering of a photon by an electron. By detecting the energy and angle of the scattered photon, the electron momentum distribution could be obtained. In the past, this experiment was done with crystal analyzers [5], which offer good energy resolution but low counting efficiency. In this paper we present a design for a TES linear array detector for X-rays ~ 100 keV that is suitable for EDXRD and Compton scattering experiments. This TES detector improves the energy resolution by an order of magnitude compared to the traditional Ge detector used in EDXRD, and can measure the full energy spectra without the time-consuming crystal rotation, which is required in the crystal analyzers.

Manuscript submitted on October 28, 2018. This work was supported by the Accelerator and Detector R&D program in Basic Energy Sciences' Scientific User Facilities (SUF) Division at the Department of Energy. This research used resources of the Advanced Photon Source and Center for Nanoscale Materials, U.S. Department of Energy Office of Science User Facilities operated for the DOE Office of Science by the Argonne National Laboratory under Contract No. DE-AC02- 06CH11357. *(Corresponding author: Antonino Miceli).*

D. Yan, O. Quaranta, A. Miceli are with Argonne National Laboratory, Argonne, IL 60439 USA, and also with Northwestern University, Evanston, IL 60208 USA (e-mail: daikangyan2013@u.northwestern.edu; amiceli@anl.gov; oquaranta@ anl.gov).

L. M. Gades, T. Guruswamy and U. M. Patel are with Argonne National Laboratory, Argonne, IL 60439 USA (e-mail: gades@anl.gov; tguruswamy@anl.gov; upatel@anl.gov)

## II. DETECTOR DESIGN

### A. Absorber geometry and energy resolution

For energy dispersive X-ray diffraction (EDXRD) experiments, segmented Ge detectors have been used to achieve the desired spatial resolution and an energy resolution of ~ 770 eV at 112 keV [6][7]. A layout similar to these segmented Ge detectors could be used on a TES linear array for the same level of spatial resolution achieved in the bulk sample. To be specific, the proposed absorber width is 0.150 mm with 0.1 mm interval trenches. A 128-pixel array has a total width of 32 mm. To reach a comparable d-spacing resolution to angle dispersive diffraction (~ 0.04 Å [8]), the detector has to improve the energy resolution by an order of magnitude.

For high-energy Compton scattering measurements carried out in synchrotron facilities, given the typical monochromatic source bandwidth and the scattering angle spread, to achieve a momentum resolution of 0.1 a.u. (atomic unit), a detector energy resolution of 80 eV is needed [5]. This energy resolution requirement is similar to that of EDXRD.

Given the geometry and energy resolution limit above, in Sec. II B we find the proper absorber length, thickness and material.

### B. Absorber material

When dealing with high-energy photons, the choosing of absorber material becomes crucial. It needs to be thick and composed of large $Z$ elements to absorb photons. In order to achieve low thermal noise, the heat capacity should be low, but cannot be too low to avoid TES saturation. Besides, the material should have good thermal conductance to achieve fast thermalization. Under such criteria, in the hard X-ray regime, Ta (tantalum), Sn (tin), Pb (lead), Bi (bismuth), Au (gold), and Cu (copper) have been used as TES absorber materials [9]-[12]. Their volume-specific heat capacity $C_v$ at 100 mK (a typical operation temperature for this kind of detector) [10]-[12], absorption coefficient $\mu$ at 100 keV, and the thickness needed to achieve ~20% quantum efficiency (QE) are listed in Table. I.

**Table 1**
Properties of absorber materials at 100 mK, 100 keV

| Material ($Z$) | $C_v$ [J·K$^{-1}$·m$^{-3}$] | $\mu$ [m$^{-1}$] | Thickness [μm] for 18% absorption |
|---|---|---|---|
| Ta (73) | 0.01 | 7162 | 27 |
| Sn (50) | 0.01 | 1226 | 162 |
| Pb (82) | 0.09 | 6298 | 32 |
| Bi (83) | 0.39 | 5592 | 35 |
| Au (79) | 7.14 | 9965 | 20 |
| Cu (29) | 9.80 | 411 | 483 |
| Ge (32) | - | 295 | 672 |

Ta, Sn, and Pb are superconductive below 100 mK, where their thermal transfer is greatly reduced because the electrons are bound into Cooper pairs that do not carry heat. Au and Cu as normal metals have high thermal conductivity, and their values have been well studied. However, the high heat capacities of Au and Cu limit the quantity that can be used in an absorber without degrading the energy resolution of the TES. To increase the QE without losing energy resolution a combination of Au and Bi can be used [13]. However, due to the lack of information on the thermal conduction properties of Bi, in this paper we consider an absorber made only of Au, with a thickness that will guarantee QE ~20%.

The lower limit of TES energy resolution can be roughly calculated [14] from

$$\Delta E = 2.35\sqrt{4 k_B T^2 C / \alpha}, \qquad (1)$$

where $k_B$ is the Boltzmann constant, and $\alpha$ is the temperature sensitivity of the TES. Assuming $\alpha = 100$, and the operating temperature $T = 100$ mK, for $C = 96.4$ pJ/K, an energy resolution of 10.7 eV can be achieved.

Based on these conditions, we design the absorber to be 4.5 mm long with 20 μm Au thickness to achieve 18% quantum efficiency at 100 keV. The absorbers have an elongated shape in order to increase the collecting area, and this can lead to position dependence in the TES response, which is discussed in Sec. II D.

### C. Other parameters

We follow the common design for the TES sensor as described in Ref. [13]. The TES has a normal resistance of $R_n$ = 9 mΩ and is biased at 15% $R_n$. Its critical temperature is 100 mK, and the heat bath is 70 mK. We also assume its current sensitivity $\beta$ has a value of 1. The shunt resistor in parallel with the TES is 0.3 mΩ. For a 150 μm ×150 μm TES, its heat capacity ($C_{TES}$) is negligible compared to that of the absorber and estimated from Ref. [13] to be 0.2 pJ/K at the TES critical temperature 100 mK. The thermal conductance between the TES and the absorber should be as large as possible to avoid heat loss during the thermalization process caused by a photon event and to minimize internal thermal fluctuation noise. A Cu patch between the TES and the absorber can be used as thermal link. Given its thermal conductivity at 300 K of $k_{300}$ = 398 W/(m·K) [15], with residual resistance ratio (RRR) of 5, typical for our Cu films, its thermal conductivity at 4.2 K can be calculated as [16]

$$\frac{k_{4.2}}{k_{300}} = \frac{RRR}{71} \qquad (2)$$

and $k_{4.2}$ = 28 W/(m·K). With a typical fabrication geometry, for which the thickness $d$ = 100 nm, width $w$ = 10 μm, and length $l$ =10 μm, the thermal conductance

$$G_{ta} = k_{4.2} \cdot \frac{d \cdot w}{l} \qquad (3)$$

is calculated to be 2.8×10$^6$ pW/K.

Figures 1 and 2 show the physical layout and thermal model of the device. Perforations are made evenly along the edges of the TES sensor and the absorber, in order to tune the thermal conductance to the thermal bath. The ratio between the size of the TES and that of the absorber gives a ratio between the respective thermal conductances ($G_{th}$ and $G_{ab}$) of $G_{tb}/G_{ab}$ = 3/61. Based on the experimental data in Ref. [13], reasonable assumptions for these quantities are $G_{tb}$ = 90 pW/K, $G_{ab}$ = 1830 pW/K.

When a photon hits a TES, its temperature and current change quickly, then relax back to the equilibrium state. Typically, an inductor is used in series with the TES to slow down the pulse response. For a microwave SQUID (superconducting quantum interference device) readout scheme, a reasonable bandwidth is around 100 kHz [17]. In order to have enough sampling points within the pulse rise time, the inductance $L$ needs be larger than 100 nH. Furthermore, $L$ cannot be so large that the electro-thermal system loses stability, which leads to an upper limit of 1000 nH. In this study, we choose $L$ = 700 nH, a value typically used in our experiments.

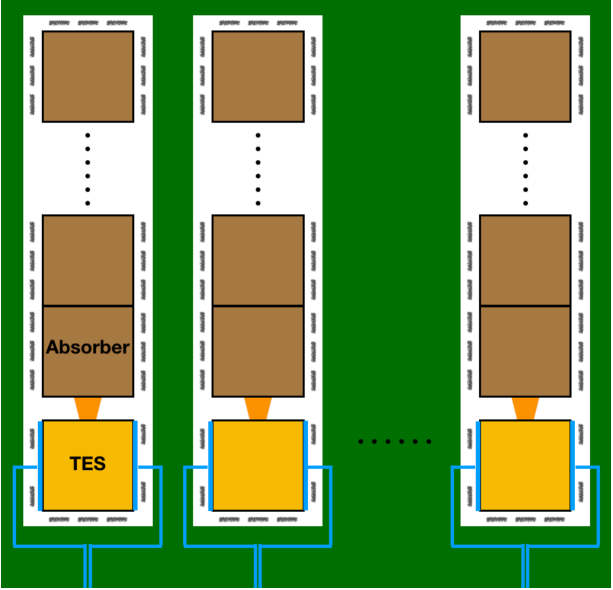

Figure 1: The layout of the TES array. The TES (yellow) is 150 μm ×150 μm, and the absorber (brown) is 150 μm × 4.5mm. These two are thermally connected via a Cu stub (orange). The spacing between each pixel is of 100 μm. This configuration offers one-dimensional spatial resolution in the horizontal direction. In order to simulate the position response, the absorber is divided to 30 squares. The devices are fabricated on a SiN membrane with 1 μm thickness (white areas), and the green area is the silicon base that connects to the cryogenic stage. Perforations are etched into the membrane evenly around the TES and absorber, so that the thermal conductance to the 70 mK heat bath is evenly distributed along the edges.

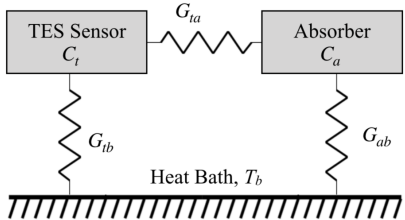

Figure 2: The device can be treated as having two thermal bodies that are the TES sensor with a heat capacity $C_t$ and the absorber with a heat capacity $C_a$. They are thermally connected through a link of thermal conductance $G_{ta}$, and are in thermal contact with the heat bath at temperature $T_b$ through links of $G_{tb}$ and $G_{ab}$, respectively.

Given the strip layout of absorber, signal position dependence must be taken into account. Here we simulate the device signal response by dividing the absorber into 30 squared units longitudinally (Fig. 1), each having a heat capacity of $C_{unit}$ = 3.2 pJ/K. The signal response can be expressed by 32 electro-thermal differential equations, taking the similar form as Ref. [18]. Given our experimental Au films RRR = 2.8, $k_{300}$ = 315 W/(m·K) [15], following Eq. (2-3), the thermal conductance between the Au absorber units $G_{abs\text{-}inter}$ is calculated to be 0.25 mW/K. When a photon with energy $E_0$ hits an absorber unit, it gives an initial temperature rise $\delta T(0) = E_0/C_{unit}$ in that unit, while in other absorber units and the TES, the temperature and current changes are zero. With this initial condition, integrating the differential equations in time series, the signal response can be calculated numerically.

### D. Simulation and results

We simulated the response of each absorber unit to a 100 keV photon. The current responses and deviations from the average are shown in Fig. 3. The position dependence can be seen clearly. Due to the differences in the pulse shapes, the calculated pulse energy will also be different, as will be shown later in this section. The pulse of the first unit deviates from the average more significantly compared with other pulses, especially after ~1 ms. In this discrete simulation, the first absorber unit is qualitatively different to the others because of its connection to the TES. This difference is more obvious when the thermal conductance between the absorber and the TES is larger.

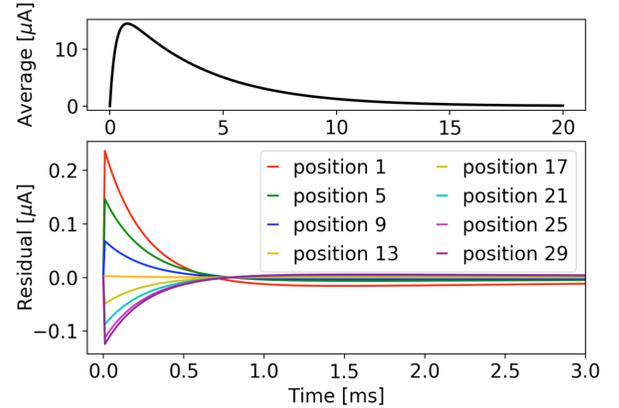

Figure 3: The average of the TES current pulse responses (upper) and the pulse deviations from the average (lower). Plots are showing signals from 8 representative incident positions that are every 4 units away from the TES.

With the device parameter settings introduced in the previous section, under the simple two-body approximation (Fig. 2), the theoretical noise of the TES is calculated and shown in Fig. 4. Given the pulse sampling rate of 100 kHz as mentioned in Sec. II C, and the pulse duration time of ~ 20 ms as shown in Fig. 3, we choose the noise spectra frequency range to be 0 Hz - 50 kHz with a bin size of 5 Hz. In practical experiments, X-ray pulses are processed with the optimal filter [14], and the average pulse shape of the 30 absorber units is usually used as the filter template. Following this routine, the device theoretical energy resolution is calculated to be 29.4 eV.

In order to evaluate the resolution broadening caused by position dependence, the 30 pulses for 100 keV photons hitting different regions are processed with the optimal filter, and the calculated energies are shown in Fig. 5. The photon hitting the unit that is closest to the TES shows a significant deviation due to its skewed pulse response, which reduces the estimated photon energy when the optimal filter is applied to it. The FWHM (full width at half maximum) of this distribution is 21.6 eV, and this gives an overall resolution of

$$\Delta E_{\text{total}} = \sqrt{29.4^2 + 21.6^2} = 36.5 \; eV.$$

This value satisfies the requirements of the target experiments. Furthermore, the degradation caused by the position dependence along the absorber can be improved by using more sophisticated pulse processing methods [19][20] or the position dependence can be used to provide spatial resolution along the strip direction [21].

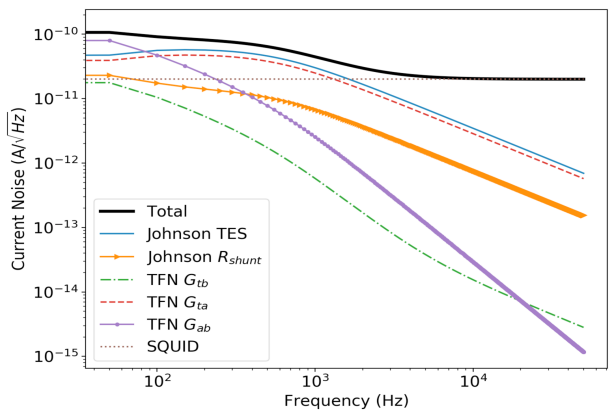

Figure 4: TES noise spectra. The noise sources taken into account are the Johnson noise of the TES and the shunt resistor, the thermal fluctuation noise (TFN) across the three thermal links from the two-body model, and the noise from the SQUID. The TFN $G_{ta}$ noise appears to have the same shape of the TES Johnson noise because the value of $G_{ta}$ is much larger than $G_{tb}$.

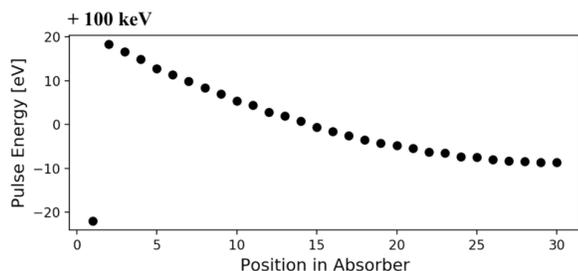

Figure 5: The energy of the photons hitting the 30 positions in the absorber calculated from the optimal filtering. The results show deviation from 100 keV. Position one is closest to the TES.

### III. CONCLUSION

We have presented a design for a TES linear array with strip absorbers to measure Compton profile and EDXRD up to 100 keV. Simulations are conducted to assess the position dependence due to the large longitudinal dimension of the absorber. We achieve an overall energy resolution of 36.5 eV, which meets the requirements of the experiments. Preliminary fabrication tests are currently taking place to produce prototype detectors.

ACKNOWLEDGMENT

This work was supported by the Accelerator and Detector R&D program in Basic Energy Sciences' Scientific User Facilities (SUF) Division at the Department of Energy. This research used resources of the Advanced Photon Source and Center for Nanoscale Materials, U.S. Department of Energy Office of Science User Facilities operated for the DOE Office of Science by the Argonne National Laboratory under Contract No. DE-AC02- 06CH11357.